\documentclass{ws-procs975x65}

\def\beq{\begin{equation}}
\def\eeq{\end{equation}}

\def\widebar{\overline}

\begin{document}

\title{Accelerating Horndeski cosmologies screening the vacuum energy}
\author{Prado Mart\'{\i}n-Moruno}

\address{Departamento de F\'{\i}sica Te\'orica I, Universidad Complutense de Madrid,\\ E-28040 Madrid,
Spain.
}

\author{Nelson J.~Nunes and Francisco S.~N.~Lobo} 

\address{Instituto de Astrof\'isica e Ci\^encias do Espa\c{c}o, Universidade de Lisboa, \\
Faculdade de Ci\^encias, Campo Grande,\\ PT1749-016 Lisboa, Portugal.
}

\begin{abstract}
In the context of Horndeski cosmologies, we consider a dynamical adjustment mechanism able to screen any value of the vacuum energy 
of the matter fields leading to a fixed de Sitter geometry. Thus, we present the most general scalar-tensor cosmological models without 
higher than second order derivatives in the field equation that have a fixed spatially flat de Sitter critical point for any kind of material 
content or vacuum energy. These models allow us to understand the current accelerated expansion of the universe as the result of the evolution 
towards the critical point when it is an attractor.
\end{abstract}

\keywords{dark energy, alternative theories of gravity, cosmology}

\bodymatter

\section{Introduction}

As is well known, the current accelerated expansion of the Universe can be described in the framework of general relativity by introducing a cosmological constant. The observed value of the energy density associated to this constant seems to be, however, much smaller than the value predicted for the energy density of the vacuum energy using quantum field theoretical considerations.

Another approach to the description of this accelerated expansion consists in considering alternative theories of gravity, the simplest theories being 4-dimensional metric theories with one extra degree of freedom. It is important to note that the new theory should not contain instabilities. In particular, the Ostrogradski instability, related with an unbounded Hamiltonian, can be avoided considering field equations without higher than second-order derivatives. The most general scalar-tensor theory with second order equations is Horndeski 
theory,\cite{Horndeski:1974wa} which is given by the Lagrangian 
\begin{eqnarray}
 \mathcal{L}_{\rm H}&=&\delta^{\alpha\beta\gamma}_{\mu\nu\sigma}\left[\kappa_1\nabla^\mu\nabla_\alpha\phi R_{\beta\gamma}{}^{\nu\sigma}
-\frac{4}{3}\kappa_{1,X}\nabla^\mu\nabla_\alpha\phi\nabla^\mu\nabla_\beta\phi\nabla^\sigma\nabla_\gamma\phi\right.\nonumber\\
&+&\left.\kappa_3\nabla_\alpha\phi\nabla^\mu\phi R_{\beta\gamma}{}^{\mu\sigma}-
4\kappa_{3,X}\nabla_\alpha\phi\nabla^\mu\phi\nabla^\mu\nabla_\beta\phi\nabla^\sigma\nabla_\gamma\phi\right]\nonumber\\
&+&\delta^{\alpha\beta}_{\mu\nu}\left[F R_{\alpha\beta}{}^{\mu\nu}-4F_{,X}\nabla^\mu\nabla_\alpha\phi\nabla^\mu\nabla_\beta\phi+
2\kappa_8\nabla_\alpha\phi\nabla^\mu\phi\nabla^\nu\nabla_\beta\phi\right]\nonumber\\
&-&3\left[2F_{,\phi}+X\kappa_8\right]+\kappa_9,
\end{eqnarray}
where
\beq
X=\nabla_\mu\phi\nabla^\mu\phi,\quad\delta^{\mu_1...\mu_n}_{\nu_1...\nu_n}=n!\delta^{\mu_1}_{[\nu_1}...\delta^{\mu_n}_{\nu_n]},\quad
F_{,X}=\kappa_{1,\phi}-\kappa_3-2X\kappa_{3,X}
\eeq  
and there are four arbitrary functions $\kappa_i=\kappa_i\left(\phi,\,X\right)$. This theory, which has been rediscovered by Deffayet 
\emph{et al.},\cite{Deffayet:2011gz} contains general relativity, Brans--Dicke and other interesting theories as particular cases.

If one is interested in considering cosmological solutions, one can restrict attention to a FLRW metric and an homogeneous scalar field $\phi(t)$ to write the minisuperspace Horndeski Lagrangian 
as\cite{Charmousis:2011bf}
\begin{equation}\label{Lmini}
L_{\rm H}=a^3\sum_{i=0}^3\left[X_i\left(\phi,\,\dot\phi\right)-\frac{k}{a^2} Y_i\left(\phi,\,\dot\phi\right)\right]\,H^i,
\end{equation}
where $L_{\rm H}=V^{-1}\int {\rm d}^3 x\,\mathcal{L}_{\rm H}$, and the functions $X_i$ and $Y_i$ are given in terms of the Horndeski functions through\cite{Charmousis:2011bf,Charmousis:2011ea}
\begin{eqnarray}
 X_0&=&-\widebar Q_{7,\phi}\dot\phi+\kappa_9, \qquad X_1=-3\,\widebar Q_7+Q_7\dot\phi,
 \\  
 X_2&=& 12F_{,X}X-12F, \quad  X_3=8\,\kappa_{1,X}\,\dot\phi^3,\\
Y_0 &=& \widebar Q_{1,\phi}\dot\phi+12\,\kappa_3\dot\phi^2-12\,F,\qquad Y_1=\widebar Q_1-Q_1\dot\phi,\qquad Y_2=Y_3=0,
\end{eqnarray}
with
\begin{eqnarray}
Q_1=\widebar Q_{1,\dot\phi}=-12\,\kappa_1, Q_7=\widebar Q_{7,\dot\phi}=6\,F_{,\phi}-3\,\dot\phi^2\kappa .
\end{eqnarray}

On the other hand, the vacuum energy also gravitates in Horndeski theory generically. Thus, if it is different from zero it could drive the dynamics of the universe at late times. As phase transitions can change the value of the vacuum energy by a finite amount, an efficient adjustment mechanism has to be able to screen different values of this vacuum energy. Weinberg pointed out in 1989\cite{Weinberg}, however, that it seems that there is no self-adjustment mechanism able to screen \emph{any} value of the vacuum energy.
Nevertheless, Charmousis \emph{et al.}\cite{Charmousis:2011bf,Charmousis:2011ea} have avoided Weinberg's conclusion by considering that the scalar field can have a nontrivial temporal dependence once screening has taken place. That is, when the geometry of the universe is described by a Minkowski solution. To understand how the field is able to self-tune, note that the modified Friedmann equation in vacuum can be written as 
\beq\label{H}
\mathcal{H}_{\rm H}(a,\,\dot a,\phi,\,\dot\phi)=-\rho_{\rm vac},
\eeq
where $\mathcal{H}_{\rm H}$ is the Hamiltonian density obtained from the minisuperspace Lagrangian (\ref{Lmini}). Thus, discontinuities in $\rho_{\rm vac}$ can be absorbed in the l.~h.~s.~of Eq.~(\ref{H}) with a continuous field $\phi$ once screening has taken place if the Hamiltonian density evaluated at the Minkowski solution depends on $\dot\phi$. As $\dot\phi$ has to be completely free to absorb any discontinuity, it cannot be restricted by the field equation, which has to be trivially satisfied when evaluated at the Minkowski solution. Moreover, any matter density non-interacting with the field appears also on the r.~h.~s.~of Eq.~(\ref{H}) and is, therefore, screened.  Thus, screening should be dynamical to allow a nontrivial cosmological evolution before complete screening has taken place. 

This self-tuning filter is, therefore, equivalent to require that Minkowski is a critical point of the dynamics.\cite{Charmousis:2011bf,Charmousis:2011ea} Actually one needs additionally that the critical point be an attractor to screen the vacuum energy when it would dominate. Nevertheless, if the universe has to approach a Minkowski solution at late times, it is not trivial to see how a late phase of accelerated expansion would be naturally described in this framework.

\section{Self-tuning to de Sitter}

It is commonly assumed that in the absence of particle physics the space is Minkowski.  One could wonder, however, whether the theory of gravity's vacuum could be described by a de Sitter space with a value  of $\Lambda$ given by the theory of 
gravity.\cite{ensayo} In this framework, an adjustment mechanism able to screen any value of the vacuum energy of matter fields leads to a de Sitter space with $\Lambda$ of the critical point of purely gravitational 
origin.\cite{modelo} Thus, the late-time accelerated expansion can be naturally described as the result of approaching the critical point, when it is an attractor.

In order to obtain the Horndeski cosmological models with a de Sitter critical point for any value of the vacuum energy and kind of material content, we restrict attention to spatially flat FLRW cosmologies, simplifying the minisuperspace Lagrangian to 
\beq\label{Lk0}
L_{\rm H}(a,\,H,\,\phi,\,\dot\phi)=a^3\sum_{i=0}^3X_i\left(\phi,\,\dot\phi\right)H^i.
\eeq
We also consider that the material content does not interact with the field; therefore, $L_{\rm m}=-a^3\rho_{\rm m}(a)$.
The modified Friedmann equation and field equation are given by
\beq\label{MFE}
\mathcal{H}_{\rm H}=\sum_{i=0}^3\left[(i-1)X_i+X_{i,\dot\phi}\dot\phi\right]H^i=\rho_{\rm m}(a),
\eeq
\beq\label{fieldE}
\mathcal{E}_{\rm H}=-\frac{{\rm d}}{{\rm d}t}\left[a^3\sum_{i=0}^3X_{i,\dot\phi}H^i\right]+a^3\sum_{i=0}^3X_{i,\phi}H^i=0,
\eeq
respectively. We apply the three self-tuning conditions summarized in the introduction, but considering that the critical point is a spatial flat de Sitter solution with a fixed value of $\Lambda$. These are: $\mathcal{H}_{{\rm H},\dot\phi}^{\rm dS}\neq0$; $\mathcal{E}_{{\rm H}}^{\rm dS}=0$, $\forall\phi$; and $\mathcal{E}_{{\rm H},\ddot a}\neq0$; with ``dS'' denoting evaluation at the de Sitter critical point. One finally obtains that the minisuperspace Lagrangian density of the models screening to de Sitter has to take the following form when evaluated at the critical 
point:\cite{modelo}
\beq\label{LdS}
\mathcal{L}_{\rm H}^{\rm dS}=3\sqrt{\Lambda}\,h(\phi)+\dot\phi\, h_{,\phi}(\phi).
\eeq
Thus, the $X_i$ functions appearing in the minisuperspace Lagrangian (\ref{Lk0}) either are linear on $\dot\phi$ or have a nonlinear dependence on $\dot\phi$ which contribution has to vanish when the Lagrangian is evaluated at the critical point.

\subsection{The magnificent seven}

The first group of models that we consider have functions $X_i$ linear on $\dot\phi$. Thus, without loss of generality, they can
be written as\cite{modelo}
\beq
\mathcal{L}_{\rm H}^{\rm ms}\left(a,\,\dot a,\,\phi,\,\dot\phi\right)=\sum_{i=0}^3X_i^{\rm ms}\left(\phi,\,\dot\phi\right)H^i,
\eeq
with
\beq\label{Xms}
X_i^{\rm ms}=3\sqrt{\Lambda} \,U_i(\phi)+\dot\phi\,W_i(\phi).
\eeq
Taking into account the constraint (\ref{LdS}), one obtains
\beq\label{const}
\sum_{i=0}^3W_i(\phi)\Lambda^{i/2}=\sum_{i=0}^3U_{i,\phi}(\phi)\Lambda^{i/2}.
\eeq
Thus, there are only seven independent functions of the field, which is the reason we denote these models the ``magnificent seven''.

The Hamiltonian density for these models can be obtained by substituting Eq.~(\ref{Xms}) into Eq.~(\ref{MFE}), which provides
\beq
\mathcal{H}_{\rm ms}=\sum_{i=0}^3\left[3(i-1)\sqrt{\Lambda}\,U_i(\phi)+i\,\dot\phi\,W_i(\phi)\right]H^i.
\eeq
This relation depends on $\dot\phi$ when evaluated at the critical point if there is at least one function $W_i(\phi)\neq0$, with $i\neq0$. The terms with $U_i(\phi)$'s and $W_0(\phi)$ do not spoil screening, although they do not self-tune by themselves. On the other hand, substituting Eqs.~(\ref{Xms}) and (\ref{const}) into Eq.~(\ref{fieldE}), the field equation can be written as
\beq
\sum_{i=0}^3 \left\{ 3\sqrt{\Lambda} \left[U_{i,\phi}(\phi)-\frac{H} {\sqrt{\Lambda}}W_i(\phi)\right]-i  \frac{\dot H}{H} W_i(\phi)  \right\}  H^i=0.
\eeq
It can be checked that it is trivially satisfied at $H_{\rm os}=\sqrt{\Lambda}$, and depends on $\dot H$ for $W_i(\phi)\neq0$, with $i\neq0$.

It must be noted that an Einstein--Hilbert term is contained within the magnificent seven. It can be explicitly written by redefining $U_2(\phi)\rightarrow U_2(\phi)-1/(8\pi G\sqrt{\Lambda})$, and it cannot screen by itself. This is not surprising because general relativity does not have a de Sitter critical point for any kind of material content. However, we can combine this term with other self-tuning models to attempt to describe a cosmology compatible with general relativity at early times.

\subsection{Their nonlinear friends}

The contribution of the $X_i$ terms with a nonlinear dependence has to vanish in the minisuperspace Lagrangian evaluated at the critical point to recover Eq.~(\ref{LdS}). Thus, writing the Lagrangian for these terms as
\beq
\mathcal{L}_{\rm H}^{\rm nl}\left(a,\,\dot a,\,\phi,\,\dot\phi\right)=\sum_{i=0}^3X_i^{\rm nl}\left(\phi,\,\dot\phi\right)H^i,
\eeq
the functions $X^{\rm nl}_i$ have to satisfy\cite{modelo}
\beq\label{const2}
\sum_{i=0}^{3} X^{\rm nl}_i\left(\phi,\,\dot\phi\right)\Lambda^{i/2}=0,
\eeq
to fulfill Eq.~(\ref{LdS}). In this case we have only three independent functions, but now they are functions of $\phi$ and $\dot\phi$, so we have much more freedom.

These models have a Hamiltonian density given by
\beq
\mathcal{H}_{\rm nl}=\sum_{i=0}^{3}\left[(i-1)X^{\rm nl}_i+\dot\phi \,X^{\rm nl}_{i,\dot\phi}\right]H^i.
\eeq
Taking into account Eq.~(\ref{const2}) and its derivatives into the Hamiltonian density evaluated at the critical point, one gets
\beq
\mathcal{H}_{\rm ms}^{\rm dS}= \sum_{i=0}^{3}i\, X^{\rm nl}_i\left(\phi,\,\dot\phi\right)\Lambda^{i/2}\neq0,
\eeq
which depends on $\dot\phi$. The field equation (\ref{fieldE}) is given by
\beq
\sum_{i=0}^{3}\left[X^{\rm nl}_{i,\phi}-3X^{\rm nl}_{i,\dot\phi}H- iX^{\rm nl}_{i,\dot\phi}\frac{\dot H}{H}-
 X^{\rm nl}_{i,\dot\phi \phi}\dot\phi-X^{\rm nl}_{i,\dot\phi\dot\phi}\ddot\phi\right] H^i=0,
\eeq
which has a nontrivial dependence on $\dot H$. Moreover, using again Eq.~(\ref{const2}) and its derivatives it can be checked that it is trivially satisfied when evaluated at the critical point.

\section{Summary and further comments}

As a first step to attempt to alleviate the cosmological constant problem, we have considered an adjustment mechanism able to screen any value of the vacuum energy. For this purpose, we have extended the concept of self-tuning in a phenomenologically useful way, assuming that complete screening leads to a de Sitter vacuum instead of Minkowski. The value of $\Lambda$ characterizing the critical point is not determined by the energy of the vacuum of the matter fields but it is of purely gravitational origin.

Following this approach we have obtained two families of cosmological scenarios that have a spatially flat de Sitter critical point for any value of the vacuum energy and kind of material content. We have referred to these models as the magnificent seven and their nonlinear friends.

The reader interested in these models can check Refs.~\citenum{lineal} and \citenum{nolineal} where the cosmology of the magnificent seven and 
their nonlinear friends have been studied. Moreover, in Ref.~\citenum{nolineal},
it has been shown that the spatially flat de Sitter critical point is an attractor for the shift-symmetric nonlinear models if
the material content of the universe satisfies the null energy condition.

\section*{Acknowledgments}
This work was partially supported by the Funda\c{c}\~{a}o para a Ci\^{e}ncia e Tecnologia (FCT) through the grants EXPL/FIS-AST/1608/2013 and UID/FIS/04434/2013. PMM acknowledges financial support from the Spanish Ministry of Economy and Competitiveness (MINECO) through the postdoctoral training contract FPDI-2013-16161, and the project FIS2014-52837-P. FSNL was supported by a FCT Research contract, with reference IF/00859/2012.

\end{document}